\documentclass[letterpaper,10pt,twocolumn]{IEEEtran}

\usepackage{amssymb,amsmath,amsfonts,bm,epsfig,graphicx,theorem,epstopdf,wasysym}
\usepackage{rotating,setspace,latexsym,epsf,color,dsfont,caption}
\usepackage{cite,subfigure,url}
\usepackage{booktabs}

\newtheorem{Theorem}{Theorem}

\newtheorem{Lemma}{Lemma}
\newtheorem{Proposition}{Proposition}
\newtheorem{Definition}{Definition}

\newenvironment{Proof}[1]{\medskip\par\noindent
{\bf Proof:\,}\,#1}{{\mbox{\,$\blacksquare$}\par}}

\newcommand{\Eb}{\mathbb{E}}
\newcommand{\Pb}{\mathbb{P}}

\newcommand{\lv}{\mathbf{1}}

\newcommand{\Ec}{\mathcal{E}}

\newcommand{\tcb}{\textcolor{black}}

\begin{document}

\title{\huge Optimal Status Update for Age of Information Minimization with an Energy Harvesting Source}

\author{Xianwen Wu$^*$,  Jing Yang$^\dagger$, and Jingxian Wu$^\ddagger$% <-this % stops a space
\thanks{$^*$Xianwen Wu is with Qualcomm Inc., San Diego, CA, USA. (email: \texttt{xianwenw@qti.qualcomm.com}).}
\thanks{$^\dagger$Jing Yang is with the School of Electrical Engineering and Computer Science, The Pennsylvania State University, University Park, PA, USA. (e-mail: \texttt{yangjing@psu.edu}).}
\thanks{$^\ddagger$Jingxian Wu is with the Department of Electrical Engineering at the University of Arkansas, Fayetteville, AR, USA (email: \texttt{wuj@uark.edu}).}
\thanks{
		This work was supported in part by the U.S. National Science Foundation (NSF) under Grants ECCS-1405403 and ECCS-1650299, and presented in part at the 2017 IEEE International Conference on Communications (ICC), Paris, France, May 2017~\cite{Yang:ICC:2017}.		}
}

\maketitle

\begin{abstract}
In this paper, we consider a scenario where an energy harvesting sensor continuously monitors a system and sends time-stamped status updates to a destination. The destination keeps track of the system status through the received updates. We use the metric Age of Information (AoI), the time that has elapsed since the last received update was generated, to measure the ``freshness" of the status information available at the destination. Our objective is to design optimal online status update policies to minimize the long-term average AoI, subject to the energy causality constraint at the sensor. We consider three scenarios, i.e., the battery size is infinite, finite, and one unit only, respectively.

For the infinite battery scenario, we adopt a best-effort uniform status update policy and show that it minimizes the long-term average AoI. For the finite battery scenario, we adopt an energy-aware adaptive status update policy, and prove that it is asymptotically optimal when the battery size goes to infinity. For the last scenario where the battery size is one, we first show that within a broadly defined class of online policies, the optimal policy should have a renewal structure. We then focus on a renewal interval, and prove that the optimal policy should have a threshold structure, i.e., if the AoI in the system is below a threshold when an energy arrival enters an empty battery, the sensor should store the energy first and then update when the AoI reaches the threshold; otherwise, it updates the status immediately. Simulation results corroborate the theoretical bounds.

\end{abstract}

\section{Introduction}
Enabled by the widespread wireless communications and the proliferation of ultra-low power sensors, ubiquitous sensing has profoundly changed almost every aspect of our daily lives. In many applications, such as environment monitoring~\cite{WSNA/Mainwaring02}, vechicle tracking~\cite{Papadimitratos09}, sensors are deployed to monitor the status of sensing objects, and communicate the status information to a monitor. To keep track of the status, it is desirable to keep the status information at the monitor as fresh as possible. However, this is often constrained by limited physical resources, such as energy and bandwidth. In order to measure the freshness of the status updates at the monitor, a metric called ``Age of Information'' (AoI) has been introduced to measure the timeliness of the status information in a network~\cite{infocom/KaulYG12}. Proposed for a node monitoring a system and sending time-stamped status updates to a destination, the metric has proved to be of fundamental importance for quantifying the freshness of information as it considers the time of generation of information in addition to network delivery delay.  Specifically, at time $t$, the AoI in the system is defined as $t-U(t)$, where $U(t)$ is the time stamp of the latest received update packet at the destination, i.e., the time at which it was acquired at the source. \tcb{AoI is fundamentally different from standard network performance metrics, such as throughput and delay. Roughly speaking, to maximize the throughput of update packets in a system, source nodes should generate as many updates as possible and push them through the network. However, heavy traffic load may congest the network and lengthen the delivery time of each update packet, which essentially increases the age of each received update, thus increasing AoI in the system. Additionally, although the age of each update is closely related to the delay it experiences in the network, update packets that get stuck in a network may become outdated after fresher update packets arrive at the destination. Thus, conventional first-come first-served (FCFS) queue management protocols are no longer desirable. Last-come first-served (LCFS) or even dropping some aged packets may become more preferable.}

There have been two main directions in the study of AoI since it was first introduced in~\cite{infocom/KaulYG12}. The first direction is to model the status updating system as a queueing system, where the update packets are generated according to a random process, and analyze the corresponding AoI under different queue management protocols. For single-server systems, all update packets from all sources are buffered in a single queue and then delivered to the destination through a single transmitter. The corresponding AoI has been analyzed in single-source single-server queues~\cite{infocom/KaulYG12}, the $M/M/1$ Last-Come First-Served (LCFS) queue with preemption in service~\cite{ciss/KaulYG12}, the $M/M/1$ First-Come First-Served (FCFS) system with multiple sources~\cite{isit/YatesK12,YatesK16}, and a multiple-source $M/M/1$ system which only keeps the latest status packet of each source in the queue~\cite{Pappas:2015:ICC}. LCFS with gamma-distributed service times and Poisson update packet arrivals is considered in \cite{isit/NajmN16}. Most recently, packet deadlines are found to improve AoI in $M/M/1$ systems in \cite{isit/KamKNWE16}, and AoI in the presence of packet delivery errors in an $M/M/1$ system is evaluated in \cite{isit/ChenH16}. A related metric, Peak Age of Information (PAoI), is introduced in \cite{isit/CostaCE14,tit/CostaCE16}, and has been studied in multi-class $M/G/1$ systems in \cite{isit/HuangM15}. Age penality function or non-linear age has been studied in \cite{Kosta:2017:nonlinear}. In systems with multiple servers, AoI has been evaluated in \cite{isit/KamKE13, isit/KamKE14,tit/KamKNE16}, and the optimality properties of a preemptive Last Generated First Served (LGFS) service discipline are identified in \cite{isit/BedewySS16, shroff_age_multi_hop}.
%For models more closely related to communication networks, AoI is evaluated in a CSMA system where multiple sources generate updates periodically and broadcast them to each other in \cite{Kaul:2011:Secon}, and in a multiple-access channel with scheduled access and slotted ALOHA-like random access protocols in \cite{Kaul:2017:MAC}. It is shown that slotted ALOHA-like random access results in longer AoI than scheduled access. Transmission scheduling in a broadcast channel with periodic update packet generation for each client is studied in \cite{Modiano:2016:BC}. It is shown that a greedy policy which always tries to update the most outdated client is optimal. Head of line (HoL) age-based scheduling algorithms have been shown to be throughput optimal in wireless networks in~\cite{Srikant:2015:ageSchedule}. Link scheduling in a multiple-source system with conflicting links for min-max peak age is studied in~\cite{He:2016:Wiopt,HeAE:ICC:2016}.
The second direction is to control the generating process of the update packets, so that the AoI is optimized.
Optimal status update policy with knowledge of the server state has been studied in~\cite{infocom/SunUYKS16}.
The relationship between AoI and the MMSE in remote estimation of a Wiener process is investigated in \cite{Sun:ISIT:2017,SunPU17}. Various source and channel coding techniques for AoI optimization have been discussed in \cite{Parag:2017:WCNC,Bhambay:2017:WCNC,Yates:2017:ISIT,Najm:2017:ISIT,ZhongYates:2016:DCC}. AoI optimization for data storage has been studied in \cite{Yates:ISIT2017:Cache}.

In parallel, energy harvesting (EH) has been well on its way to becoming a game-changing technology in the field of autonomous wireless networked systems. The notion of acquiring energy from nature to power wireless transmitters is intriguing since careful transmission scheduling can render extended or even perpetual operation of the network \cite{Yang_tcom, kaya_tcom, ozel11}. Optimal transmission scheduling for throughput and delay optimization has been studied under both infinite battery setting~\cite{Yang_tcom,Ho:TSP:2012,ozel11} and finite battery setting~\cite{kaya_tcom,ozel_finite_tcom,Ozel0U13,Srivastava:2013:BPL,Dor:JSAC:2016}.
With signal processing related performance metrics, such as detection delay and estimation error, optimal sensing scheduling policies have been developed to optimize the sensing performances of EH sensor networks~\cite{Kar:2006:DNA,Lai:quickest,Yang:jsac:2015,Yang:jsac:2016}. 

Age of Information in EH wireless networks is in its infancy with only a few recent works that investigate various status update policies under an energy harvesting setting, in very specific setups \cite{isit/Yates15,ita/BacinogluCU15,BacinogluU17,Arafa:AoI:2017}. It has been shown, in \cite{isit/Yates15} that in this setting, with knowledge of the system state, updates should be submitted only when the server is free to avoid queueing delay. Moreover, a greedy policy that submits a fresh update as the system becomes idle is shown to be inefficient; a {\it lazy} update policy that introduces inter-update delays is better. The optimal update policy remains open even in this setting. In \cite{ita/BacinogluCU15}, under the assumption that a status update packet can be generated and served (transmitted) instantly, the authors investigate optimal offline and online policies. The optimal offline policy is to equalize the inter-update delays as much as possible, subject to the energy constraint imposed by the energy harvesting source. The online problem is cast as a Markov decision process in a discrete-time setting, and solved through dynamic programming. Other threshold type status update policies have been studied in \cite{BacinogluU17} and shown to be optimal under certain conditions. An offline policy to miminimize AoI in a two-hop relay channel is studied in \cite{Arafa:AoI:2017}.

%Under an EH setting, \cite{isit/Yates15,ita/BacinogluCU15} investigate several status update polices assuming the battery at the energy harvesting sensor is sufficiently large. It has been shown in \cite{isit/Yates15} that with knowledge of the system state, updates should be submitted only when the server is free to avoid queueing delay. Moreover, a greedy policy that submits a fresh update as the system becomes idle is shown to be inefficient; a {\it lazy} update policy that introduces inter-update delays is better. The optimal update policy remains open in this setting. In \cite{ita/BacinogluCU15}, under the assumption that a status update packet can be generated and served (transmitted) instantly, the authors investigate optimal offline and online policies. The optimal offline policy is to equalize the inter-update delays as much as possible, subject to the energy constraint imposed by the energy harvesting source. The online problem is cast as a Markov Decision Process in a discrete-time setting, and solved through dynamic programming. Although it is analytically intractable, the optimal policy is shown to have a threshold structure. I.e., with real-time knowledge of the energy arrival profile and its statistics, the source sends a status update if the {\it expected} AoI is above certain threshold, given it has sufficient energy. 

In this paper, we investigate optimal {\it online} status update policies for an energy harvesting source with various battery sizes in a continuous-time setting. Similar to \cite{ita/BacinogluCU15,BacinogluU17}, we assume a status update packet can be generated by the source at any time and transmitted to a destination instantly, given sufficient energy is available at the source. We assume that the energy unit is normalized so that each status update requires one unit of energy. This energy unit represents the cost of both measuring and transmitting a status packet.
We assume energy arrives at the sensor according to a Poisson process, and the sensor only has causal information of the energy arrival profile in addition to the parameter of the Poisson process. Our objective is then to determine the sequence of update instants so that the long-term average AoI at the destination is minimized, subject to the energy causality constraint at the source.

We first study the properties of the time-average AoI as a function of inter-update delays, and establish a connection between this problem and the optimal sensing problem studied in~\cite{Yang:jsac:2016}. This motivates us to adopt the (asymptotically) optimal sensing policies in \cite{Yang:jsac:2016} for AoI minimization, namely, a best-effort uniform status update policy for the infinite battery case, and an energy-aware adaptive status update policy for the finite battery case. Since the AoI function does not have all the properties required to establish the optimality of those policies in~\cite{Yang:jsac:2016}, we revise the proofs accordingly to re-establish their (asymptotic) optimality. We then study a special case where the battery size is one unit, and propose a threshold based status update policy, i.e., if the AoI in the system is below a threshold when an energy enters an empty battery, the sensor should store the energy and hold status update until the AoI reaches the threshold; otherwise, it consumes the energy to update the status immediately. Through rigorous stochastic analysis, we show that within a broadly defined class of online policies, this threshold based status update policy is optimal.

% We first identify a performance limit on the long-term time average sensing performance of the system. Motivated by the structure of the performance limit, we propose a best-effort uniform sensing policy, and prove that it achieves the limit asymptotically, thus it is optimal. When the battery size is finite, we aim to investigate the impact of finite battery size on the sensing performance, and bring the sensing performance as close to that of the system with infinite battery as possible. We propose an energy-aware adaptive sensing scheduling policy, which dynamically chooses the next sensing epoch based on the battery level at the current sensing epoch, and show that it is asymptotically optimal as the battery size increases. The convergence rate is also explicitly characterized.

\section{System Model and Problem Formulation} \label{sec:model}
Consider a scenario where an energy harvesting sensor continuously monitors a system and sends time-stamped status updates to a destination. The destination keeps track of the system status through the received updates. We use the metric Age of Information (AoI) to measure the ``freshness" of the status information available at the destination.

\tcb{In a typical wireless sensor network, the measurement and radio frequency transmission processes consume power in the range of $1\sim 100$  $mW$, and take a few seconds or less. Meanwhile, typical output power under average conditions for different EH technologies, such as indoor solar cells, piezoelectric cells and wireless power transfer, ranges from below 100 $\mu W$ to hundreds of $\mu W$~\cite{Raghunathan:2005,piezo:2009,Bhatti:2016:EHW}. Roughly speaking, it takes about a few minutes or more for an EH node to charge its battery in order to perform one status update. }

Therefore, in the following, we assume that the time used to collect and transmit a status update is negligible compared with the time scale of inter-update delays, i.e., given sufficient energy is available at the source, a status update can be generated by the source at any time and transmitted to the destination instantly. In this case, a status update is transmitted immediately after it is generated to avoid unnecessary queueing delay. \tcb{We assume the channel between the source and the destination is noiseless, thus the transmitted update always gets delivered successfully. We leave the more general setting where the channel is noisy and updates may be corrupted and unrecognizable at the destination as our future work.}

We assume that the energy unit is normalized so that each status update requires one unit of energy. This energy unit represents the cost of both measuring and transmitting a status update. \tcb{In the following, we assume a Poisson energy arrival process to make the theoretical analysis easier to track. We will relax this assumption in the simulations in Section~\ref{sec:simulation}.} Assume energy arrives at the sensor according to a Poisson process with parameter $\lambda$. Hence, energy units arrive at discrete time instants $t_1,t_2,\ldots$. We assume $\lambda=1$ throughout this paper for ease of exposition. The sensor is equipped with a battery with capacity $B$, $B\geq 1$. When $B=\infty$, it corresponds to the infinite battery case.

% a typical power consumption profile for a sensor node within a WSN intended for environmental monitoring. Most of the time the nodes are set to a low power state (sleep mode in Figure 1), being periodically wake up to acquire the desired environmental parameters, which are then conveyed to the network host. The slow variation of environmental conditions allows programming sleep times of tens of minutes, while the power-hungry measurement and radio frequency (RF) transmission processes take a few seconds or less

A status update policy is denoted as $\pi:=\{S_n\}_{n=1}^\infty$, where $S_n$ is the $n$-th update epoch. We assume $S_0=0$, i.e., the system updates its status information right before time zero. Denote the inter-update delays as $X_n\triangleq S_n-S_{n-1}$, for $n=1,2,\ldots$. Then, we have $S_n=\sum_{i=1}^nX_i$.

Define $A(X_n)$ as the total amount of energy harvested in $[S_{n-1},S_n)$, and $E(S^-_n)$ as the energy level of the sensor right before the scheduled updating epoch $S_n$. For a clear exposition of the paper, we assume the system has one unit amount of energy before it updates at time zero, and after that, the battery becomes empty, i.e., 
\begin{align}
E(S_0^-)=1.\label{eqn:energy_initial}
\end{align}
Then, under any feasible status update policy, the energy queue evolves as follows
\begin{align}
E(S^-_{n})&=\min\{E(S^-_{n-1})-1+A(X_{n}),B\}, \label{eqn:energy_queue}\\
E(S_n^-)&\geq 1, \label{eqn:energy_constraint}
\end{align}
for $n=1,2,\ldots$. Equation (\ref{eqn:energy_constraint}) corresponds to the energy causality constraint in the system.
Based on the Poisson arrival process assumption, $A(X_{n})$ is an independent Poisson random variable with parameter $X_{n}$. 

\begin{figure}[t]
\centering
\includegraphics[width=3in]{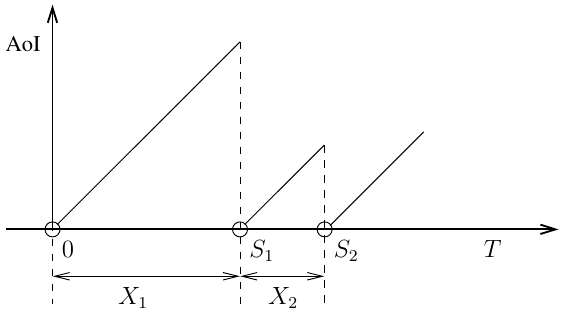}
\caption{AoI as a function of $T$. Circles represent status update instants. }\label{fig:AoI}
\end{figure}

Under any feasible status update policy, the AoI as a function of time is shown in Fig.~\ref{fig:AoI}. We use $N(T)$ to denote the number of status updates generated over $(0,T]$. Define $R(T)$ as the total ``reward", i.e., age of information experienced by the system over $[0,T]$. Then,
\begin{align}\label{defn:R(T)}
R(T)&=\frac{\sum_{i=1}^{N(T)}X_i^2+ (T-S_{N(T)})^2}{2},
\end{align}
and the time average AoI over the duration $[0,T]$ can be expressed as $R(T)/T$.

Our objective is to determine the sequence of update epochs $S_1,S_2,\ldots$, so that the time average AoI at the FC is minimized, subject to the energy causality constraint. We focus on a set of {\it online} policies $\Pi$ in which the information available for determining the updating epoch $S_n$ includes the updating history $\{S_i\}_{i=0}^{n-1}$, the energy arrival profile over $[0,S_n)$, as well as the energy harvesting statistics (i.e., $\lambda$ in this scenario).
The optimization problem can \tcb{be} formulated as
\begin{eqnarray}\label{eqn:opt}
\underset{\pi\in\Pi}{\min} & &\limsup_{T\rightarrow +\infty} \Eb\left[\frac{R(T)}{T}\right]\\
\mbox{s.t. } & & (\ref{eqn:energy_initial})-(\ref{eqn:energy_constraint}),\nonumber
\end{eqnarray}
where the expectation in the objective function is taken over all possible energy harvesting sample paths. This problem does not admit \tcb{an} MDP formulation in general, and it is extremely challenging to explicitly identify the optimal solution.

\section{Optimal Status Updating when $B$ is large}\label{sec:policy}
In \cite{Yang:jsac:2016}, we studied an optimal sensing scheduling problem. Our objective was to strategically select the sensing epochs, so that the long-term average sensing performance can be optimized. We assumed that the sensing performance over $[0,T]$ can be expressed as $\sum_{i=1}^{N(T)}f(X_i)+f(T-N(T))$, where $X_i$ is the $i$-th inter-sensing delay. Under the assumption that 1) $f(x)$ is convex and monotonically increasing in $x$; 2) $f(x)/x$ is increasing in $x$; and 3) $f(x)/x$ is upper bounded by a positive constant, we proposed two sensing policies, for the infinite and finite battery cases, respectively, and proved their (asymptotic) optimality.

We note that the AoI minimization problem can be treated as a particularized case of the optimal sensing scheduling problem studied in  \cite{Yang:jsac:2016}, by replacing the general sensing performance metric with AoI. Thus, for this particular case, $f(x)=x^2/2$. We note that this function exhibits the first two properties required to establish the optimality of the proposed sensing scheduling policies in \cite{Yang:jsac:2016}. However, the last condition i.e.,  $f(x)/x$ is upper bounded by a positive constant, does not hold, due to the fact that $f(x)/x=x/2$ and it is unbounded. Therefore, the optimality of the policies proposed in \cite{Yang:jsac:2016} need to be carefully examined. In the following, we will utilize the specific form of the AoI function to bypass the last condition and reaffirm the optimality of the policies.

For the completeness of this paper, in this section, we adapt the major results and policies in \cite{Yang:jsac:2016} for the AoI minimization setup. We leave out the proofs that do not reply on the third assumption, and provide necessary new proofs only. \tcb{We will start with the infinite battery case, and investigate its performance lower bound and the corresponding bound-achieving status updating policy. The policy is shown to have a uniform updating structure. With insights drawn from the infinite battery case, we will then study the finite battery case. We will develop an energy-aware status updating policy by modifying the uniform updating policy, and show that as the battery size increases, it approaches the uniform updating policy, thus it is asymptotically optimal. }

\subsection{Status Update with Infinite Battery}  \label{sec:infinite}
When the battery size is infinite, no energy overflow will happen. Thus, the maximum achievable long-term average status update rate is one update per unit time. If we drop the energy causality constraint, and replace it with this long-term average status update rate constraint, we obtain a lower bound on the long-term average AoI, \tcb{which is 1/2.} 
This lower bound corresponds to a uniform status update policy which updates once per unit time. However, it may become infeasible when the energy causality constraint is imposed. Thus, we propose the following policy to ensure the status update policy is always feasible.

\begin{Definition}[Best-effort Uniform Status Update Policy]
The sensor is scheduled to update the status at $s_n=n$, $n=1,2,\ldots$. The sensor performs the task at $s_n$ if $E(s^-_n)\geq 1$; Otherwise, the sensor keeps silent until the next scheduled status update epoch.
\end{Definition}

Here we use $s_n$ to denote the $n$-th {\it scheduled} status update epoch, which is in general different from the $n$-th {\it actual} status update epoch $S_n$ since some of the scheduled status update epochs may be infeasible.

\begin{Theorem}\label{thm:myopic}
The best-effort uniform status update policy is optimal when the battery size is infinite, i.e.,
\begin{align*}
\limsup_{T\rightarrow +\infty}\frac{R(T)}{T} & =  \frac{1}{2}\quad a.s..
\end{align*}
%where $d_n$ is the duration between the actual status update epochs $l_n$ and $l_{n-1}$.
\end{Theorem}

The proof of Theorem~\ref{thm:myopic} is provided in Appendix~\ref{appx:thm2}.
Intuitively, when the battery size is infinite, the fluctuation in the energy harvesting process can be averaged out when $T$ is sufficiently large, thus the uniform status update policy can be achieved asymptotically. %However, with finite battery, the lower bound may not be achieved under the best-effort uniform updating policy, since energy overflow is inevitable in this situation, which in turn results in more frequent infeasible sensing epochs due to battery outage.

\subsection{Status Update with Finite Battery}  \label{sec:finite}
In order to minimize the long-term average AoI when the battery size is finite, intuitively, the status update policy should try to prevent any battery overflow, as wasted energy leads to performance degradation. Meanwhile, the properties of AoI require the status update rate to be as uniform as possible in time. Those two objectives are not aligned with each other, thus, the optimal status update policy should strike a balance between them.

In the following, we propose an energy-aware adaptive status update policy, which adaptively changes the update rate based on the instantaneous battery level. When the battery level is high, the sensor updates more frequently in order to prevent battery overflow; When the battery level is low, the sensor updates less frequently to avoid infeasible status update epochs. Meanwhile, the update rate does not vary significantly in time in order to control the increase of time-average AoI caused by the jittering updating epochs.

\begin{Definition}[Energy-aware Adaptive Status Update Policy] Assume $B>1$. The adaptive status update policy defines status update epochs $s_n$ recursively as follows
\begin{align}\label{eqn:adaptive_sensing}
s_n&=s_{n-1}+\left\{\begin{array}{cl}\vspace{0.05in}
\frac{1}{1-\beta}, & E(s^-_{n-1})<\frac{B}{2}\\
\vspace{0.05in}
1, & E(s^-_{n-1})=\frac{B}{2}\\
\frac{1}{1+\beta}, & E(s^-_{n-1})>\frac{B}{2}
\end{array}\right.,
\end{align}
where $s_0=0$, $E(s^-_0)=1$, and
%\begin{align}
$\beta:=\frac{k\log B}{B}$,
%\end{align}
with $k$ being a positive number such that $0<\beta<1$.
The sensor samples and updates the status at $s_n$ if $E(s^-_n)\geq 1$; Otherwise, the sensor keeps silent until the next scheduled status update epoch.
\end{Definition}

As $B\rightarrow \infty$, we have $\beta\rightarrow 0$ for any fixed $k$, i.e., the adaptive status update policy converges to the best-effort uniform status update policy as battery size increases. Thus, \tcb{we expect that the long-term average AoI under the adaptive status update policy converges to that under the best-effort uniform status update policy, which is 1/2, as the battery size approaches infinity.}

\tcb{Let $f$ and $g$ be two functions defined on some subset of the real numbers. Denote $f(x)=O(g(x))$ if and only if $\lim_{x\rightarrow 0}\frac{|f(x)|}{|g(x)|}\leq M$, where $M$ is a positive constant. Then, the asymptotic optimality of the adaptive status update policy is described in the following theorem. }
%The asymptotic optimality of the adaptive status update policy, and the corresponding convergence rate are described in the following two theorems.

%\begin{Theorem}\label{thm:probability}
%Under the adaptive status update policy, the proportion of infeasible status update epochs scales in $O\left(\frac{2^{k+1}k(\log B)^2}{B^{k+1}}\right)$, and the average amount of wasted energy per unit time scales in $O\left(\frac{2^{k+1}k(\log B)^2}{B^{k+1}}\right)$.
%\end{Theorem}

%Theorem~\ref{thm:probability} indicates that when $B$ is sufficiently large, both upper bounds of the battery outage and overflow probabilities decrease monotonically as $k$ increase. As the battery size $B$ increases, the upper bounds of those two probabilities decrease and eventually approaches zero. \tcb{Thus, the proposed policy is asymptotically equivalent to a uniform sensing policy, similar to the best-effort uniform sensing policy for the infinite battery case.}

\begin{Theorem}\label{thm:utility}
Under the adaptive status update policy, the gap between the long-term average AoI and its lower bound $1/2$ scales in $O\left(\frac{2^{k+1}k(\log B)^2}{B^{k+1}}+\left(\frac{\log B}{B}\right)^2\right)$ almost surely.
\end{Theorem}
Theorem~\ref{thm:utility} implies that as battery size $B$ increases, the long-term average AoI under the adaptive status update policy approaches 1/2, which is the lower bound on the long-term average AoI in a system with infinite battery. Thus, it is asymptotically optimal.
%Compared to the bounds in Theorem~\ref{thm:probability}, the bound in Theorem \ref{thm:utility} has an extra term $\left(\frac{\log B}{B}\right)^2$.  For a sufficiently large $B$, the bound is dominated by the first term when $k$ is small, and it is dominated by the second term when $k$ is large. Thus, it may not monotonically decrease as $k$ increase, which is consistent with the fact that the time average AoI is not only related to the battery outage and overflow probabilities, but also depends on the durations between consecutive status update epochs.
The proof of Theorem \ref{thm:utility} is provided in Appendix~\ref{appx:utility}.

\section{A Special Case: $B=1$}
In the previous section, we investigate the optimal and asymptotically optimal status update policies when battery size $B$ is infinite, and finite but sufficiently large, respectively. However, when the battery size is so small that the asymptotics cannot kick in, those policies may not perform very well. This motivates us to investigate other status update policies when battery size $B$ is small. One extreme case for this scenario is when $B=1$, i.e., the battery can only store the energy for one status update operation. In this case, the battery only has two states: empty, or full. When it is empty, obviously, any status update should not be scheduled. When one unit amount of energy arrives, the battery jumps to the other state, and it then \tcb{needs} to decide when to spend the energy for status update. Denote $\Gamma_i$ as time duration between $S_{i-1}$ and the first energy arrival time after $S^-_{i-1}$. Then, we have the following observations

\begin{Lemma}\label{lemma:exp}
When $B=1$, under any feasible online policy, we must have $X_i\geq \Gamma_i$, $\forall i$, and $\Gamma_i$s are independent and identically distributed (i.i.d.) random variables, with common distribution $\exp(1)$.
\end{Lemma}
Lemma~\ref{lemma:exp} is based on the energy causality constraint, and the memoryless property of the inter-arrival times of the Poisson energy arrivals.

As defined in Sec.~\ref{sec:model}, the policy space $\Pi$ includes all of the {\it online} policies which make the status updating decision based on up-to-date updating history and energy arrival profile, as well as the energy harvesting statistics. In other words, $X_i$ is a function of $\Gamma_i$, among other variables. 

In order to facilitate the analysis, in the following, we focus on a special class of online policies, termed as uniformly bounded policies.

\begin{Definition}[Uniformly bounded policy]\label{dfn:uniform}
For an online policy with $\{(X_i,\Gamma_i)\}_{i=1}^{\infty}$, if $\forall i$, $\Gamma_i\leq X_i$, and there exists a function $g(\Gamma_i)$ such that $X_i \leq g(\Gamma_i)$, and the second moment of $g(\Gamma_i)$ is finite, then this policy is a uniformly bounded policy.
%A uniformly bounded policy is a policy under which $\Gamma_i\leq X_i(\Gamma_i) \leq g(\Gamma_i)$, $\forall  \Gamma_i\in \Rb_+$, $\forall i\in \Nb_+$, where $g(x)$ is a function with finite moments, .
\end{Definition}

%Denote the set of uniformly bounded policies as $\Pi'$, 

\begin{Theorem}\label{thm:comparison}
Any uniformly bounded policy is sub-optimal to a renewal policy, i.e., a policy under which the updating epochs $\{S_i\}_{i=1}^{\infty}$ form a renewal process. Besides, under the renewal policy, $X_i$ only depends on $\Gamma_i$.
\end{Theorem}
The proof is provided in Appendix~\ref{appx:comparison}. Our approach involves two steps of averaging. The first step of averaging is in the space of status update sample paths under a given uniformly bounded policy. For each fixed $i$ and $\tau$, we group all of the sample paths with $\Gamma_i=\tau$ , and obtain the corresponding average inter-update delay $X_i(\tau)$. This step essentially averages out all factors that may affect $X_i$ other than $\Gamma_i$. The second step is to do an averaging in the temporal domain. For each fixed $\tau$, we form a sophisticated linear combination of involved $X_i(\tau)$s, and use it as the inter-update delay under the new policy. Such a policy is a renewal policy, it is always feasible, and each renewal interval only depends on $\tau$. Through rigorous stochastic analysis, we prove that the new renewal policy always outperforms the original policy in terms of time-average AoI. 

In the following, we will focus on renewal policies, and show that the threshold structure of the optimal renewal policy in the following theorem.
\begin{Theorem}\label{thm:renewal}
In the class of renewal policies, the optimal policy has a threshold structure, i.e., $X_i$ equals a constant $\tau_0$ if $\Gamma_i\leq \tau_0$; otherwise $X_i$ equals $\Gamma_i$. Here $\tau_0=0.9012$, and the corresponding long-term average AoI equals $\tau_0$.
\end{Theorem}
The proof of Theorem~\ref{thm:renewal} is provided in Appendix~\ref{appx:renewal}. 

%The proof is sketched as follows: We note that the long-term average AoI under a renewal policy can be expressed in the form of $\frac{\Eb[X^2]}{2\Eb[X]}$, where $X$ is a random variable with the same distribution of $X_i$. The objective is then to identify the optimal $X$ as a function of $\tau$, which follows the same as $\Gamma_i$. We first establish a necessary condition for $X(\tau)$ to be optimal, i.e., if we add a small perturbation on $X(\tau)$ without violating the energy causality condition, the objective function can only increase. Such a necessary condition indicates that  $X(\tau)$ must have the threshold structure described in Theorem~\ref{thm:renewal}.

Theorem~\ref{thm:renewal} indicates the optimality of the following threshold-based status update policy.
       \begin{Definition}[Threshold-based Status Update Policy]
%%       The threshold-based status update policy keeps tracking the last status update time.
       When an energy unit enters an empty battery, the sensor performs a status update immediately if the AoI at the FC is greater than a threshold $\tau_0$; Otherwise, it holds its operation until the AoI is exactly equal to $\tau_0$.
       \end{Definition}

%The long-term average AoI under this policy can be analytically characterized based on the memoryless property of the exponentially distributed inter-arrival times of energy units. %We summarize the result in the following theorem.
%
%\begin{Theorem}\label{thm:threshold}
%Under the threshold-based status update policy, the long-term average AoI is $h(\tau_0)\triangleq \frac{2\tau_0 e^{-\tau_0}+2e^{-\tau_0}+\tau_0^2}{2(e^{-\tau_0}+\tau_0)}$.
%\end{Theorem}
%The proof of Theorem~\ref{thm:threshold} can be found in Appendix~\ref{appx:threshold}. Besides, by taking derivative of $h(\tau_0)$, we can show that $h(\tau_0)$ is first decreasing, then increasing in $\tau_0$. Therefore, the optimal $\tau_0$ corresponds to the point where $h'(\tau_0)=0$. Solving the equation, we have $\tau^*_0=0.9012$, and the corresponding long-term average AoI is 0.9012.

%Comparing with the status update policies proposed in Section~\ref{sec:policy}, we have the following observation.
%We can further show that the threshold-based status update policy is optimal, as indicated by the following theorem.
%\begin{Theorem}\label{thm:comparison}
%Let $\Pi'\subseteq\Pi$ be a class of online policies under which the updating epochs $\{l_i\}_{i=1}^{\infty}$ form a renewal process. When $B=1$, the threshold-based status update policy minimizes the long-term average AoI among all policies in $\Pi'$.
%\end{Theorem}

\section{Simulation Results} \label{sec:simulation}
\tcb{We consider a wearable device powered by a piezoelectric energy harvester which harvests energy at an average rate 10 $\mu W$ per second. The device sends update packets periodically to a monitoring device, such as a cell phone, through low-power transmission technologies, such as Bluetooth or Zigbee. Each update consumes energy 1 $mJ$ and lasts for one second. We normalize the unit energy to be 1 $mJ$, and the unit time to be $10^2$ seconds. Therefore, the EH rate is equivalent to one unit energy per unit time. We will first evaluate the proposed policies with a Poission EH process, and then study them with a first-order Markov process.}

\begin{figure}[t]\centering
\includegraphics[width=3.4in]{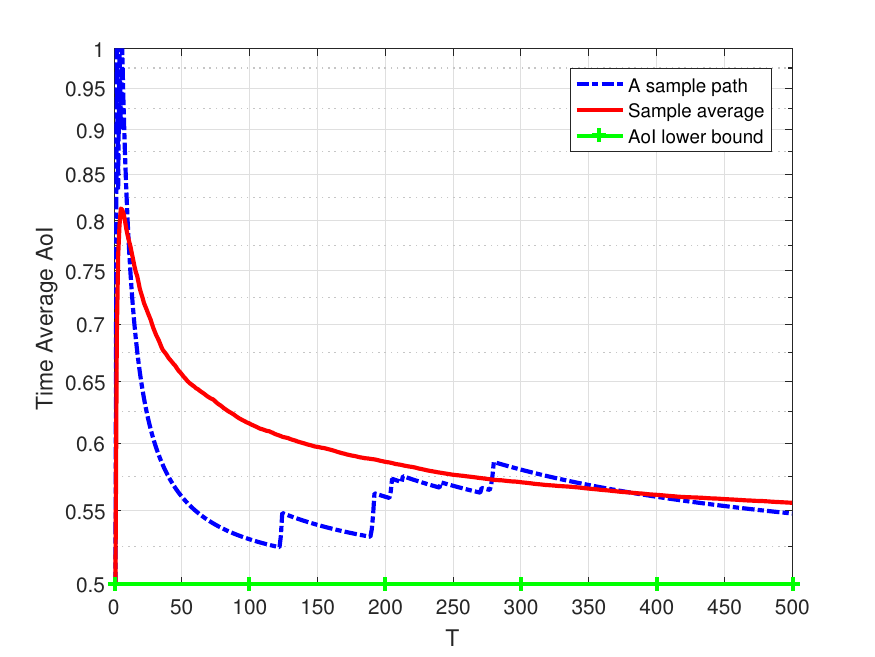}
\vspace{-0.1in}
\caption{Time average AoI under best-effort uniform status update policy.}\label{fig:AoI_uniform}
\end{figure}

First, we fix the battery size $B=\infty$. We generate sample paths for the Poisson EH process over $500 \times 10^2$ $s$, and perform status updating according to the best-effort uniform status update policy. The time-average AoI as a function of $T$ is shown in Fig. \ref{fig:AoI_uniform}. We plot one sample path of the time-average AoI and the sample average over $10^3$ sample paths in the figure. We observe that both curves gradually approach the lower bound $0.5\times 10^2$ $s$ as $T$ increases. When $T = 500\times 10^2$ $s$, there is only a very small difference between the simulation results and the analytical lower bound. The results indicate that the proposed best-effort uniform status update policy is optimal.

\begin{figure}[t]
\begin{center}
\includegraphics[width=3.4in]{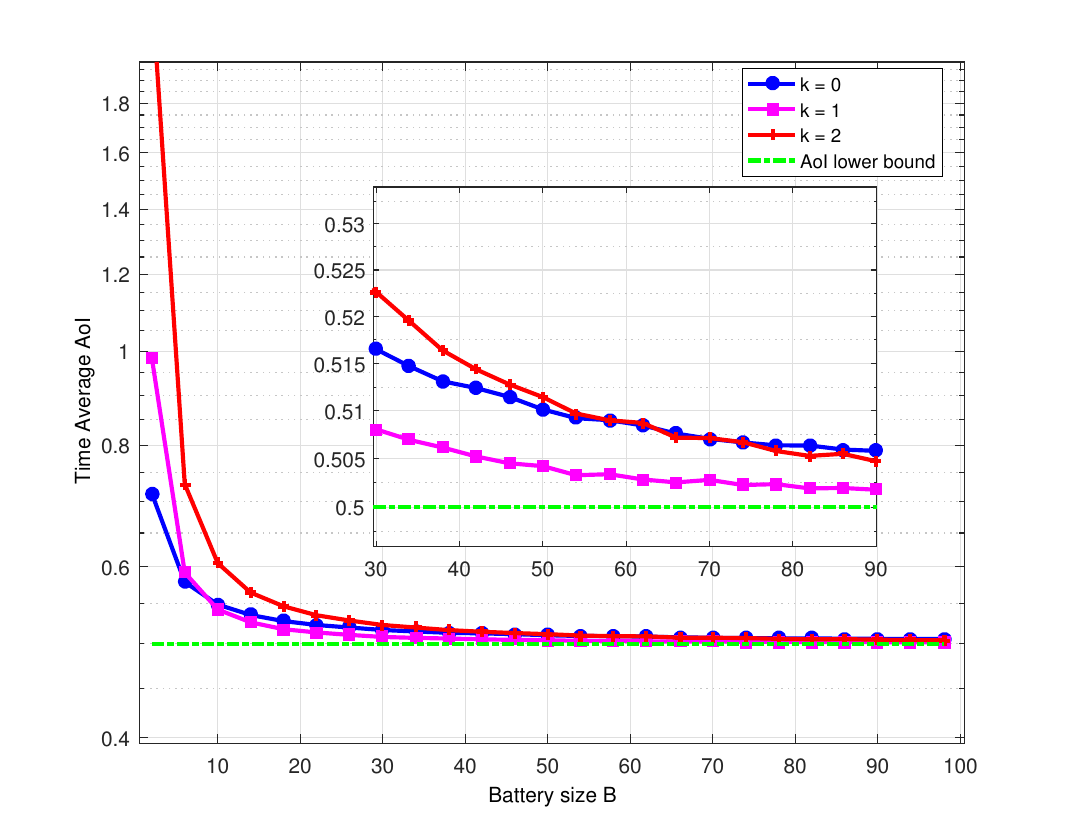}
\end{center}
\vspace{-0.1in}
\caption{Time average AoI under adaptive status update policy.}
\label{fig:utility}
\end{figure}

Next, we study the time average AoI under the adaptive status update policy with finite battery sizes. We fix $T=10^5$ time units and plot the average AoI over $10^3$ sample paths in Fig.~\ref{fig:utility}. We note that for each fixed $k$, the gap between the time average AoI and the lower bound monotonically decreases as $B$ increases, which is consistent with Theorem~\ref{thm:utility}.

\begin{figure}[t]
\begin{center}
\includegraphics[width=3.4in]{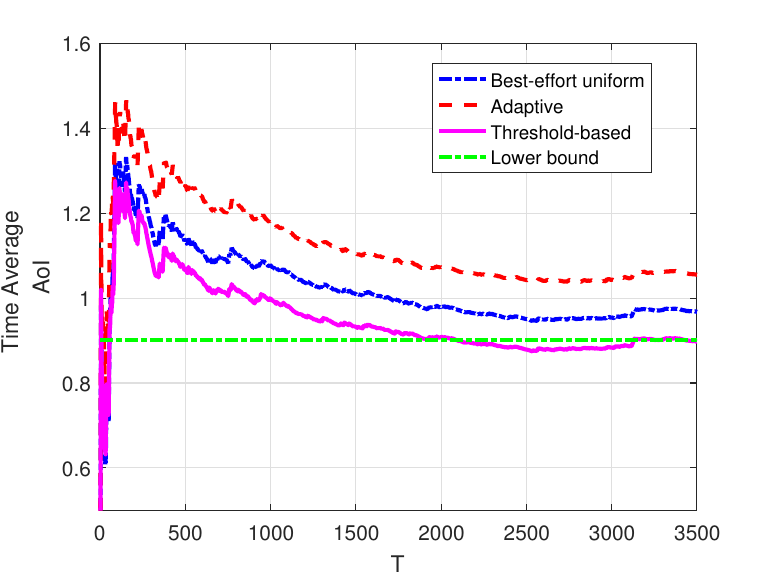}
\end{center}
\vspace{-0.1in}
\caption{Performance comparison when $B=1$: A sample path.}
\label{fig:compare}
\end{figure}

\begin{figure}[t]
\begin{center}
\includegraphics[width=3.4in]{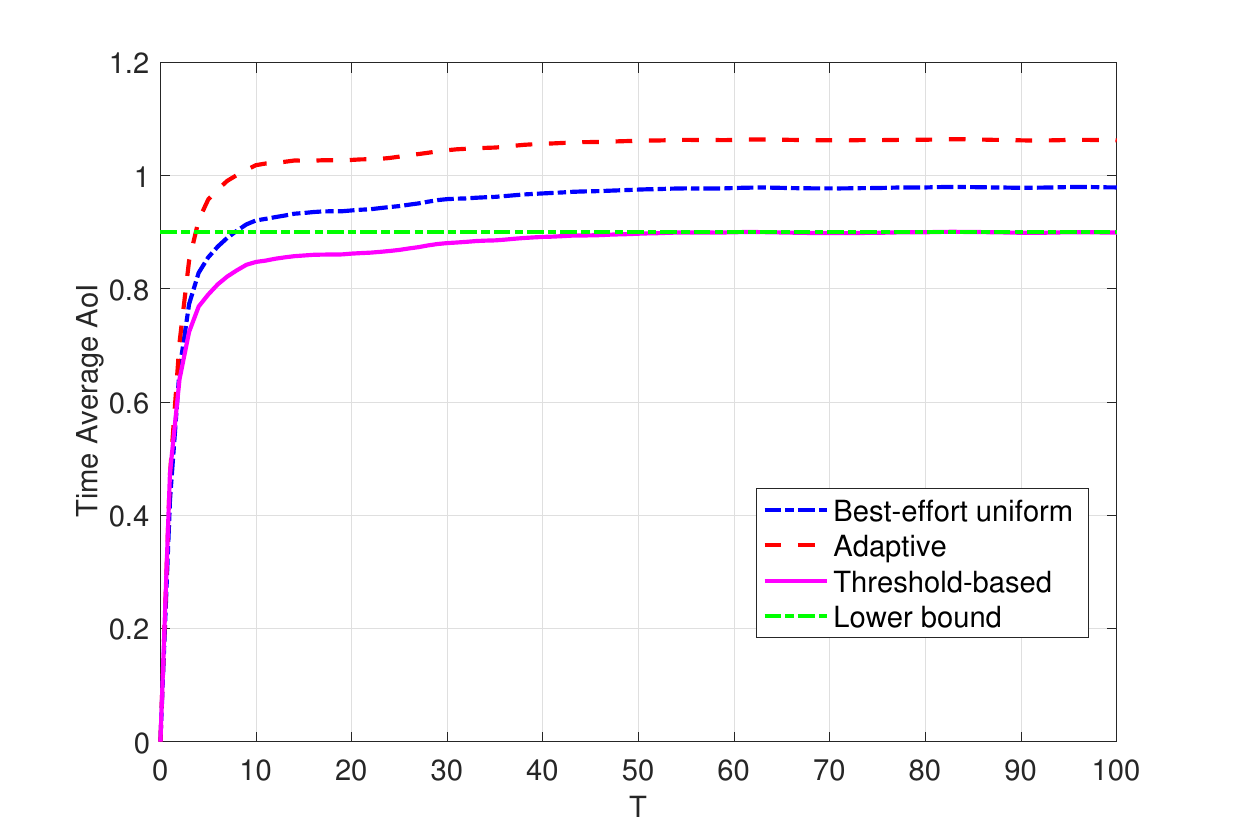}
\end{center}
\vspace{-0.1in}
\caption{Performance comparison when $B=1$: Sample-path average.}
\label{fig:compare2}
\end{figure}

Then, we compare the performances of the three policies for $B=1$. For a fair comparison, we optimize the parameters for the best-effort uniform status update policy and the adaptive status update policy numerically before we perform the comparison. We note that the optimal update rate for the best-effort uniform policy is once every $0.4302\times 10^2$ seconds. We also modify the adaptive status update policy to make it applicable for the case $B=1$. Specifically, we schedule the next update $\frac{1}{1+\beta}\times 10^2$ seconds away if the battery level is full right before the current update; otherwise, we schedule it in time $\frac{1}{1-\beta}\times 10^2$ seconds. We numerically search for the optimal value of $\beta$, and it turns out that when $\beta=-0.1450$, the time-average AoI is minimized.
This is opposite to the case when $B$ is large but finite. Although it is a bit counter intuitive, it is due to the memoryless property of the inter-arrival times of a Poisson process, i.e., the expected waiting time for the next energy arrival keeps unchanged after current scheduled update epoch, regardless of its feasibility. If $B=1$ at current scheduled update epoch, the battery will become empty immediately after it updates the status, and the AoI will then linearly grow from zero; If $B=0$, the AoI has a positive value already, and will grow with the same rate. Thus, in order to balance the inter-update delays to minimize the time average AoI, the system should be more aggressive to update if the current scheduled update is infeasible.
We then generate a sample path and plot the time average AoI as a function of time units $T$ under each policy, as shown in Fig.~\ref{fig:compare}. The corresponding sample-path average over $10^3$ sample paths is plotted in Fig.~\ref{fig:compare2}.  As we expect, for both scenarios, the threshold based updating policy outperforms the other two policies, and approaches its limit as $T$ gets sufficiently large. 

\tcb{Last, we evaluate the performance of the proposed algorithms with Markovian EH processes, which has been typically assumed in the literature~\cite{Seyedi:2008:Markov}. Specifically, we model the EH process as a stationary first-order discrete-time Markov chain with two states, namely, ON and OFF, and the length of each time slot will be decided later.  We assume that at the end of each time slot the device will go from OFF to ON with probability $p_0$, and from ON to OFF with probability $p_1$. Furthermore, we assume that the energy harvesting device will harvest 1 $mJ$ energy in one time slot if the state is ON, and does not harvest any energy in the OFF state. Thus, the steady state probability that the device is in the OFF and ON states are $\frac{p_1}{p_0+p_1}$ and $\frac{p_0}{p_0+p_1}$, respectively, and the average EH rate is thus equal to $\frac{p_0}{p_0+p_1}$ $mJ$ per time slot. To make the EH rate consistent with that under the Poisson setting for comparison, we normalize the length of each time slot as $\frac{p_0}{p_0+p_1}\times 10^2$ second, so that the EH rate for the Markov process is always equal to 1 $mJ/10^2s$. The values of $p_0$ and $p_1$ controls the burstiness of the EH process. When $p_0=p_1=1$, the EH process becomes a uniform process, and when $p_0, p_1$ are very small, the EH process becomes very bursty, i.e., it may be ON and harvest energy in consecutive time slots for many time slots, and then switch to OFF and be inactive for a long period of time. Intuitively, the bursty EH process will results in a larger AoI than the uniform EH process.}

\tcb{In the following, we consider various values of $p_0$ and $p_1$, and evaluate the time-average AoI under the proposed policies for different values of $B$. Specifically, we generate $10^3$ EH sample paths over $10^5\times 10^2$ seconds under each setting. For each sample path, we run the policies and track the time-average AoI. We then average over the sample paths to get the average AoI, and summarize them in Table~\ref{table:markov}.  When $B=\infty$, we run the best-effort uniform status updating policy which updates once every unit time if it has sufficient energy. As we note in Table~\ref{table:markov}, when $p_0=p_1=1.0$, the resulted average AoI is exactly equal to the lower bound, i.e., 0.5 unit time. This is because no battery outage would happen with to the uniform EH process. For the rest cases, the average AoI are close to 0.5. This is because the probability of battery outage approaches zero when $T\rightarrow\infty$ for $B=\infty$. The AoI monotonically increases as the EH process becomes more bursty, which is consistent with our intuition. When $B=10$, we perform the energy-aware adaptive status updating with parameter $k=1$. As we observe in the table, the average AoI is close to $0.5$ when $p_0=p_1=1.0$, and it exhibits the same monotonicity in $p_0,p_1$ as for the $B=\infty$ case. When $B=1$, we choose the threshold based updating with threshold $\tau_0=0.9012$. Again, we note that the AoI monotonically decreases as $p_0,p_1$ increase, and it exactly equals $0.5$ when $p_0=p_1=1.0$. This is because when the EH process is uniform, the AoI in the system is always equal to one when an energy unit arrives. Since it is above the threshold, the sensor will perform an update immediately, which leads to the uniform updating. } 

\tcb{The simulation results indicate that although we assume a Poisson EH process for the ease of theoretical analysis, such an assumption may not be critical for the optimality of the proposed policies, especially for the cases that $B=\infty$ and $B$ is finite but large. Theoretically characterizing the performance of these policies with more general EH processes is one of our future steps.}
\begin{table}[t]
\centering
\begin{tabular}{c|ccc}
  \toprule[1.5pt]
  Setting & $B=\infty$ & $B=10$& $B=1$ \\
  \midrule
  $p_0=p_1=0.1$ & 0.5212 & 1.2069& 2.2291\\
  $p_0=p_1=0.3$ &  0.5039 & 0.5627&0.9855\\
  $p_0=p_1=0.5$ & 0.5018&0.5224 & 0.6991 \\
  $p_0=p_1=0.7$ & 0.5009 &0.5152 & 0.5761\\
  $p_0=p_1=1.0$ &0.5000 &0.5009 & 0.5000\\
  \bottomrule[1.5pt]
\end{tabular}
\caption{Average AoI for Markovian EH processes.}\label{table:markov}
\end{table}
 
\section{Conclusions} \label{sec:conclusion}
In this paper, we investigated optimal status update policies for an energy harvesting source equipped with a battery. We considered three different cases, namely, the battery size is infinite, finite but large, and one unit. We proposed three different status updating policies for those cases, and established their optimality through theoretical analysis. We also evaluated the performances of the proposed policies through simulation results.

We point out that the (asymptotically) optimal status update polices for the infinite battery and finite battery cases are closedly related to our earlier work \cite{Yang:jsac:2016}. In \cite{Yang:jsac:2016}, we have studied an optimal sensing problem where the objective is to optimize the long-term average sensing performance. We assume that the sensing performance was measured by a general function of the inter-sensing delays. Examples include the MMSE in reconstructing a wide-sense stationary random process. We observe that the average AoI as a function of the inter-update delays can be treated as a particularized case of that general function. Such inherent connection between AoI minimization and a general sensing performance optimization implies that AoI as a metric of information freshness does have deep connections with other performance metrics in remote sensing/estimation systems. Unveiling the intricate relationship between AoI and other remote estimation related metrics is one of our future directions.

\appendix
\subsection{Proof of Theorem~\ref{thm:myopic}}\label{appx:thm2}
To prove Theorem~\ref{thm:myopic}, it suffices to show that
$\limsup_{T\rightarrow +\infty} \frac{R(T)}{T} \leq   1/2$ almost surely.

The uniform best-effort status update policy partitions the time axis into slots, each with length $1$.
Let $E(n^-)$ be the energy level of the sensor right before the {\em scheduled} status update at time $n$. 
Based on $E(n^-)$, we can group the time slots into intervals labeled as $v_1, u_1, \ldots, v_k, u_k, \ldots$, where $u_i$ corresponds to the $i$-th interval that begins with $E(n^-)=0$ for some $n$ and ends when $E(n^-)$ becomes positive as $n$ increases; $v_i$ corresponds to the $i$-th interval that begins with $E(n^-)>0$  for some $n$ and ends when $E(n^-)$ becomes zero as $n$ increases. Note that we assume one unit energy is available at time $0$, i.e., $E(0^-)=1$. 

We note that $E(n^-)$ jumps from zero to some positive value at the end of $u_i$, due to random energy arrivals over the last time slot in $u_i$. Based on assumption that the energy arrivals follow a Poisson process, the length of $u_i$ follows an independent geometric distribution where
\begin{align}\label{eqn:geometric}
  \Pb\left[ u_i = k \right] = e^{-(k-1)} ( 1 - e^{-1} ), \quad k = 1, 2\ldots
\end{align}
With a bit abuse of notation, in equation (\ref{eqn:geometric}), and in the following proofs, we use $u_i$ to denote the length of the interval labeled as $u_i$; similarly for $v_i$.

Over the interval labeled as $v_i$, all of the scheduled status update epochs are feasible, except for the last one bounding $v_i$.  
Considering the duration bounded by the first and last feasible status update epochs over $v_i$, the aggregated AoI equals $(v_i-1)f(1)$, where $f(x)=x^2/2$. Since all of the scheduled status updating epochs over $u_i$ are infeasible except for the last one (which is also the first feasible status update epoch over $v_{i+1}$), the aggregated AoI over the duration bounded by the last feasible status update epoch over $v_i$ and the first feasible status update epoch over $v_{i+1}$ is $f(u_i+1)$. 

Let $K(T)$ be the number of $u_i$s over $[0,T]$. Then the number of $v_i$s over $[0,T]$ is either $K(T)$ or $K(T)+1$, depending on whether time $T-1$ is a feasible update epoch or not. Therefore,
\begin{align}
&  \limsup_{T\rightarrow +\infty} \frac{R(T)}{T}\nonumber\\
%& =  \limsup_{T\rightarrow +\infty} \frac{\sum_{i=1}^{K(T)} \left[ (v_i - 1) f(1) + f(u_i+1) \right] }{T}\nonumber\\
& = \limsup_{T\rightarrow +\infty} \frac{\sum_{i=1}^{K(T)} f(u_i+1)}{T}
+ \frac{T-\sum_{i=0}^{K(T)} ( u_i+1) }{T}f\left(1\right) \label{eqn:segment}\\
& = \limsup_{T\rightarrow +\infty} \frac{\sum_{i=1}^{K(T)}  (u_i+1)^2}{2T}  + \frac{1}{2}  - \frac{\sum_{i=1}^{K(T)}u_i}{2T}  -\frac{K(T)}{2T} \nonumber\\
& = \limsup_{T\rightarrow +\infty} \left(\hspace{-0.02in}\frac{\sum_{i=1}^{K_T}  u_i^2}{2K(T)}+\frac{\sum_{i=1}^{K(T)}  u_i}{K(T)}+1\right)\frac{K(T)}{T} +\frac{1}{2} ,\label{eqn:AoI}
\end{align}
where (\ref{eqn:segment}) follows from the definition of $v_i$ and $u_i$, (\ref{eqn:AoI}) follows from the results that $K(T)/T\rightarrow 0$ and $\sum_{i=1}^{K(T)}u_i/T\rightarrow 0$ almost surely, as proved in the proof of Theorem~1 in \cite{Yang:jsac:2016}. Since $u_i$'s are i.i.d. geometric random variables, $\frac{\sum_{i=1}^{K(T)}  u_i}{K(T)}$ and $\frac{\sum_{i=1}^{K(T)}  u_i^2}{K(T)}$ converges to the first and second moments of the geometric distribution specified in~(\ref{eqn:geometric}), which are finite constants. Therefore, we have (\ref{eqn:AoI}) converges to 1/2 almost surely.

\subsection{Proof of Theorem~\ref{thm:utility}}\label{appx:utility}
Consider the first $n$ scheduled status update epochs under the proposed adaptive status update policy for a sample path of the energy harvesting process. Let $n_+$ denote the number of intervals between two scheduled status updating epochs with duration $\frac{1}{1-\beta}$, $n_-$ be that with duration $\frac{1}{1+\beta}$, and $n_0$ be that with duration $1$. Let $\bar{n}$ be the number of status updating epochs the battery overflows, and $\underline{n}$ be the number of infeasible status update epochs.
Then, the $n$-th scheduled status update epoch happens at time $T_n:=\frac{n_+}{1-\beta}+n_0+\frac{n_-}{1+\beta}$. Let $A_n^+$ be the total amount of energy wasted. Then,
\begin{align}
E(T_n^-)=(A(T_n)-A_n^+)-(n-\underline{n}),
\end{align}
where $A(T_n)$ is a Poisson random variable with parameter $T_n$. Dividing both sides by $n$ and taking the limit as $n$ goes to $+\infty$, we have\begin{align*}
\lim_{n\rightarrow\infty}\frac{E(T_n^-)}{n}&=\lim_{n\rightarrow\infty}\frac{A(T_n)}{T_n}\cdot\frac{T_n}{n}\hspace{-0.02in}-\hspace{-0.02in}\lim_{n\rightarrow\infty}\frac{A_n^+}{n}\hspace{-0.02in}-\hspace{-0.02in}\left(1\hspace{-0.02in}-\hspace{-0.02in}\lim_{n\rightarrow\infty}\frac{\underline{n}}{n}\right).
\end{align*}
According to Theorem 3 in \cite{Yang:jsac:2016}, for almost every sample path, 
\begin{align}
\lim_{n\rightarrow\infty}\frac{A_n^+}{n}&=O\left(\frac{2^{k+1}k(\log B)^2}{B^{k+1}}\right),\\
\lim_{n\rightarrow\infty}\frac{\underline{n}}{n}&=O\left(\frac{2^{k+1}k(\log B)^2}{B^{k+1}}\right).\label{eqn:outage}
\end{align}
Combining with the fact that $\lim_{n\rightarrow\infty}\frac{E(T_n^-)}{n}=0$ and $\lim_{n\rightarrow\infty}\frac{A(T_n)}{T_n}=1$, we have
\begin{align}\label{eqn:T_n}
\lim_{n\rightarrow\infty}\frac{T_n}{n}&=1+O\left(\frac{2^{k+1}k(\log B)^2}{B^{k+1}}\right).
\end{align}
Based on Taylor expansion and (\ref{eqn:T_n}), we have
\begin{align*}
&\lim_{n\rightarrow\infty}\frac{n_+f\left(\frac{1}{1-\beta}\right)+n_0f(1)+n_-f\left(\frac{1}{1+\beta}\right)}{T_n}\\
%&=\frac{n_+}{n}\left(f(1)+f'(1)\left(\frac{1}{1-\beta}-1\right)+O\left(\left(\frac{\beta}{1-\beta}\right)^2\right)\right)\\
%&\quad \frac{n_0}{n}f(1)+\frac{n_-}{n}\left(f(1)+f'(1)\left(\frac{1}{1+\beta}-1\right)+O\left(\left(\frac{\beta}{1-\beta}\right)^2\right)\right)\\
%&=f(1)+\left(\frac{\frac{n_+}{1-\beta}+n_0+\frac{n}{1-\beta}}{n}\right)\cdot f'(1)+O\left(\left(\frac{\log B}{B}\right)^2\right)\\
&=f(1)+O\left(\frac{2^{k+1}k(\log B)^2}{B^{k+1}}+\left(\frac{\log B}{B}\right)^2\right).
\end{align*}
On the other hand, due to the existence of infeasible status update epochs, we have
\begin{align}
&\lim_{n\rightarrow\infty}\frac{\sum_{n}f(X_n)-\left[n_+f\left(\frac{1}{1-\beta}\right)+n_0f(1)+n_-f\left(\frac{1}{1+\beta}\right)\right]}{T_n}\nonumber\\
&\leq \hspace{-0.03in}\lim_{n\rightarrow\infty}\hspace{-0.03in}\frac{\sum_{n: X_n> \frac{1}{1-\beta}} f(X_n)}{T_n}\hspace{-0.03in}=\hspace{-0.02in}\lim_{n\rightarrow\infty}\hspace{-0.03in}\frac{\sum_{n: X_n> \frac{1}{1-\beta}} X_n^2}{2\tilde{n}}\frac{\tilde{n}}{T_n},\label{eqn:AoI2}
\end{align}
where the inequality in (\ref{eqn:AoI2}) follows from the fact that $X_n$ differs from the delay between two consecutive scheduled status update epochs only when battery outage happens, and $\tilde{n}$ denote the number of $X_n$'s with $X_n> \frac{1}{1-\beta}$.

We note that for all $X_n\geq \frac{1}{1-\beta}$, $\tilde{X}_n\triangleq X_n(1-\beta)$ follows a geometric distribution with parameter $p\triangleq 1 - e^{-\frac{1}{1-\beta}}$, and its second moment is $\frac{2-p}{p^2}$. Then,
\begin{align}
&\lim_{n\rightarrow\infty}\frac{\sum_{n: X_n> \frac{1}{1-\beta}} X_n^2}{2\tilde{n}}= \lim_{n\rightarrow\infty}\frac{\sum_{n: X_n> \frac{1}{1-\beta}} (\tilde{X}_n)^2}{2(1-\beta)^2 \tilde{n}}\\
&=\frac{\Eb[\tilde{X}_n^2]}{(1-\beta)^2}= \frac{2-p}{2p^2(1-\beta)^2}. %\qquad\mbox{a.s.}
\end{align}
Meanwhile, we have $\lim_{n\rightarrow\infty}\frac{\tilde{n}}{T_n}\leq \lim_{n\rightarrow\infty}\frac{\underline{n}}{T_n}$. Combining with (\ref{eqn:outage})(\ref{eqn:AoI2}), we have
\begin{align*}
\lim_{n\rightarrow\infty}\frac{\sum_{n}f(X_n)}{T_n}=\frac{1}{2}+O\left(\frac{2^{k+1}k(\log B)^2}{B^{k+1}}+\left(\frac{\log B}{B}\right)^2\right).
\end{align*}

\subsection{Proof of Theorem~\ref{thm:comparison}}\label{appx:comparison}
Let $\{S_i\}_{i=1}^{\infty}$ be the status update epochs under a uniformly bounded policy, and $\{X_i\}_{i=1}^{\infty}$ be the corresponding inter-update delays. Based on the definition of $R(T)$ in (\ref{defn:R(T)}), we have
\begin{align}
\frac{R(S_{N(T)})}{T}\leq \frac{R(T)}{T}\leq \frac{R(S_{N(T)+1})}{T}.
\end{align} 
Thus, 
\begin{align}
\Eb \left[ \frac{R(T)}{T}\right]&\geq  \Eb\left[\frac{R(S_{N(T)})}{T}\right]\\
&=\Eb \left[\frac{R(S_{N(T)+1})}{T}\right]-\Eb \left[ \frac{X^2_{N(T)+1}}{2T}\right].
%&\geq \Eb\left[\frac{R(S_{N(T)+1})}{S_{N(T)+1}}\right] -\Eb\left[ \frac{X^2_{N(T)+1}}{2T}\right]
\end{align}
We aim to show that 1) $\Eb \left[\frac{X^2_{N(T)+1}}{2T}\right]\rightarrow 0$, and 2) $\Eb\left[\frac{R(S_{N(T)+1})}{T}\right]$ is suboptimal to a renewal policy. In the following, we will show them separately.

\subsubsection{ $\Eb \left[\frac{X^2_{N(T)+1}}{2T}\right]\rightarrow 0$}
First, we denote $F_n(t)$ as the cumulative distribution function (cdf) of $S_n$ under the uniformly bound policy, and $N(t)$ be the total number of updates over $(0, t]$. Then, we have
\begin{align}\label{eqn:cdf}
\Eb[N(t)]&=\sum_{n=0}^\infty F_n(t).
\end{align}

Next, we note that
\begin{align}
&\Eb \left[X_{n+1}^2\lv_{S_{n+1}>T}\mid S_n=t\right]\\
&=\Eb \left[X_{n+1}^2\lv_{X_{n+1}>T-t}\mid S_n=t\right]\\
&\leq\Eb_{\Gamma_{n+1}}\left[g^2(\Gamma_{n+1})\lv_{g(\Gamma_{n+1})>T-t}\mid S_n=t\right]\label{eqn:g_bound}\\
&=\Eb_{\Gamma_{n+1}}\left[g^2(\Gamma_{n+1})\lv_{g(\Gamma_{n+1})>T-t}\right]\triangleq G(T-t),\label{eqn:Gamma}
\end{align}
where (\ref{eqn:g_bound}) follows from the definition of uniformly bounded policy, and (\ref{eqn:Gamma}) follows from the memoryless property of the inter-arrival time of a Poisson process.

Therefore, by fist conditioning on the last update epoch prior to (or at) time $t$, we have 
\begin{align}
&\Eb\left[ X^2_{N(T)+1}\right]=\sum_{n=0}^\infty \int_0^T \Eb \left[X_{n+1}^2\lv_{S_{n+1}>T}\mid S_n=t\right] dF_n(t)\nonumber\\
&\leq \int_0^T G(T-t) d\left(\sum_{n=0}^\infty F_n(t) \right)\label{eqn:equation}\\
&=\int_0^T G(T-t) d \Eb[N(t)],\label{eqn:deri}
\end{align}
where (\ref{eqn:equation}) follows from (\ref{eqn:Gamma}),  and (\ref{eqn:deri}) follows from (\ref{eqn:cdf}).

For any fixed $0\leq\Delta\leq T $, we have
\begin{align}
&\frac{1}{T}\int_0^T G(T-t) d \Eb[N(t)]\nonumber\\
&=\frac{1}{T}\int_0^{T-\Delta} \hspace{-0.05in}G(T-t) d \Eb[N(t)]+\frac{1}{T}\int_{T-\Delta}^T\hspace{-0.05in} G(T-t) d \Eb[N(t)]\nonumber\\
&\leq G(\Delta)\frac{\Eb[N(T\hspace{-0.03in}-\hspace{-0.03in}\Delta)]}{T}\hspace{-0.03in}+\hspace{-0.03in}G(0)\frac{\Eb[N(T)]\hspace{-0.03in}-\hspace{-0.03in}\Eb[N(T\hspace{-0.03in}-\hspace{-0.03in}\Delta)]}{T}, \label{eqn:segment}
\end{align}
where (\ref{eqn:segment}) follows from the fact that $G(t)$ is monotonically decreasing in $t$.

Note that $A(t)$ is defined as the total number of energy arrivals over $[0,t]$, which upper bounds the total number of status updates over $(0,t]$, i.e., $N(t)$, due to energy causality constraint. Thus,
\begin{align}
&N(T)-N(T-\Delta)\nonumber\\
&= N(T)-(N(T-\Delta)+1)+1\\
&\leq A(S_{N(T)})-A(S_{N(T-\Delta)+1})+1\\
&\leq A(T)-A(T-\Delta)+1
\end{align}
under each status update sample path.
Plugging in (\ref{eqn:segment}) and letting $T\rightarrow \infty$, we have
\begin{align}
&\lim_{T\rightarrow \infty}\frac{1}{T}\int_0^T G(T-t) d \Eb[N(t)]\nonumber\\
&\leq \lim_{T\rightarrow \infty} G(\Delta)\frac{T-\Delta}{T}+G(0)\frac{\Delta+1}{T}\\
&=G(\Delta),\label{eqn:G_Delta}
\end{align}
where (\ref{eqn:G_Delta}) holds for any $\Delta\geq 0$, due to the assumption in Definition~\ref{dfn:uniform} that $\Eb[g^2(\Gamma_i)]$ is bounded. Since $\lim_{\Delta\rightarrow\infty}G(\Delta)=0$, we have $\Eb \left[\frac{X^2_{N(T)+1}}{2T}\right]\rightarrow 0$ as $T\rightarrow \infty$.

\subsubsection{$\Eb\left[\frac{R(S_{N(T)+1})}{T}\right]$ is sub-optimal to a renewal policy}
For any given uniformly bounded policy, we will construct a renewal policy as follows: For all of the status update sample paths under the given uniformly bounded policy, we will group those with $S_{i-1}\leq T$ based on the value of $\Gamma_i$, and find the corresponding average inter-update delay $X_i$. Specifically, we define
\begin{align}
\hat{X}^T_i(\tau)&\triangleq\Eb[X_i |S_{i-1}\leq T, \Gamma_i=\tau].
\end{align}
Since each $X_i\geq \Gamma_i$ under the given policy, we have $\hat{X}^T_i(\tau)\geq \tau$, and it depends only on $\tau$. 
Besides, we have the following observation:
\begin{Lemma}\label{lemma:factorization}
For any fixed $T> 0$, 
\begin{align*}
\Eb[X_i \lv\{i\leq N(T)+1\}]&=\Eb_\tau[\hat{X}^T_i(\tau)]\cdot\Eb[\lv\{i\leq N(T)+1\}].
\end{align*}
\end{Lemma}
\begin{Proof}
Based on the property of conditional expectation, we have
\begin{align}
\hat{X}^T_i(\tau)&=\frac{\Eb[X_i \lv\{S_{i-1}\leq T\}|\Gamma_i=\tau]}{\Pb[S_{i-1}\leq T|\Gamma_i=\tau]}\label{eqn:indicator}\\
&=\frac{\Eb[X_i \lv\{i\leq N(T)+1\}|\Gamma_i=\tau]}{\Eb[\lv\{i\leq N(T)+1\}]},\label{eqn:hatX}
\end{align}
where $\lv\{\Ec\}$ in (\ref{eqn:indicator}) is an indicator function, which takes value 1 if event $\Ec$ is true; otherwise, it equals 0. 
Equation (\ref{eqn:hatX}) follows from the fact that events $S_{i-1}\leq T$ and $i\leq N(T)+1$ are equivalent, and
\begin{align*}
\Pb[S_{i-1}\leq T|\Gamma_i=\tau]&=\Pb[S_{i-1}\leq T]=\Eb[\lv\{i\leq N(T)+1\}].
\end{align*}
The Lemma is proved after taking expectation of both sides of (\ref{eqn:hatX}) with respect to $\tau$.
\end{Proof}

Next, we will construct a renewal policy based on the definition of $\hat{X}^T_i(\tau)$. Define 
\begin{align}
\rho_i^T&\triangleq \frac{\Eb[\lv\{i\leq N(T)+1\}]}{\Eb[N(T)+1]},\label{dfn:rho}\\
\bar{X}_T(\tau)&\triangleq \sum_{i=1}^\infty \rho_i^T \hat{X}^T_i(\tau).\label{dfn:barX}
\end{align}
Then, we have the following observations.
\begin{Proposition}\label{prop:pmf}
For any fixed $T\geq 0$, $\{\rho_i^T\}_{i=1}^{\infty}$ is a valid distribution. 
\end{Proposition}
This proposition can be proved based on the facts that $\rho_i^T\geq 0$, and $\sum_{i=1}^{\infty}\rho_i^T=1$.

\begin{Proposition}\label{prop:renewal}
For any fixed $T\geq 0$, $\bar{X}_T(\tau)\geq \tau$, and it depends on $\tau$ only.
\end{Proposition}
This proposition is due to $\hat{X}^T_i(\tau)\geq \tau$, and it depends only on $\tau$, as well as Proposition~\ref{prop:pmf}.
Proposition~\ref{prop:renewal} indicates that if we define a status update policy such that the corresponding inter-update delay is determined by the delay between the last status update epoch and the first energy arrival time after that according to $\bar{X}_T(\tau)$, then, the corresponding policy always satisfies the energy causality constraint, and the inter-update delays over $[0,T]$ are independent and identically distributed, thus it is a renewal policy over $[0,T]$. 

With a little abuse of notation, in the following, we use $\tau$ to denote a random variable that has the same distribution as $\Gamma_i$.

%\begin{align}\label{eqn:barX_mean}
%m_1(\bar{X}_T):=\Eb[\bar{X}_T]&=\sum_{i=1}^\infty \rho^i_T \Eb_\tau[\hat{X}_T^i(\tau)]
%\end{align}

\begin{Lemma}\label{lemma:first_moment}
For any fixed $T> 0$, 
\begin{align}\label{eqn:first_moment}
\frac{\Eb[N(T)+1]}{T}&\geq \frac{1}{\Eb_\tau[\bar{X}_T(\tau)]}=\frac{1}{\sum_{i=1}^\infty \rho_i^T \Eb_\tau[\hat{X}^T_i(\tau)]}.
\end{align}
\end{Lemma}
\begin{Proof}
First, we note that $S_{N(T)+1}:=\sum_{i=1}^{N(T)+1}X_i\geq T$. Thus, we have
\begin{align}
1&\leq \Eb\left[\frac{S_{N(T)+1}}{T} \right]\\
&= \frac{1}{T}\Eb\left[\sum_{i=1}^{\infty} X_i \cdot \lv\{i\leq N(T)+1\} \right]\\
&=\frac{1}{T}\sum_{i=1}^{\infty}\Eb\left[ X_i \cdot \lv\{i\leq N(T)+1\} \right]\label{eqn:sumE}\\
&=\frac{1}{T} \sum_{i=1}^{\infty}\Eb_{\tau}[\hat{ X}_T^i(\tau)]\cdot\Eb\left[ \lv\{i\leq N(T)+1\} \right]\label{eqn:factorization}\\
&=\left(\sum_{i=1}^{\infty}\rho_i^T\Eb_\tau[\hat{ X}_T^i(\tau)]  \right) \cdot\frac{\Eb[N(T)+1]}{T}\label{eqn:rho}\\
&=\Eb_\tau[\bar{X}_T(\tau)]\cdot\frac{\Eb[N(T)+1]}{T}\label{eqn:barX}
\end{align}
where we switch the order of summation and expectation in (\ref{eqn:sumE}) since $X_i\geq 0$, (\ref{eqn:factorization}) follows from Lemma~\ref{lemma:factorization}, (\ref{eqn:rho}) follows from the definitions of $\rho_i^T$ in (\ref{dfn:rho}), and (\ref{eqn:barX}) follows from the definition of $\bar{X}_T$ in (\ref{dfn:barX}). Dividing $\Eb_\tau[\bar{X}_T(\tau)]$ on both sides of (\ref{eqn:barX}), we have (\ref{eqn:first_moment}) proved.
\end{Proof}

\begin{Lemma}\label{lemma:cauchy}
Under the uniformly bounded policy, we have
\begin{align*}%\label{eqn:cauchy2}
&\left(\Eb[X_i \cdot \lv\{i\leq N(T)+1\}|\Gamma_i=\tau]\right)^2\nonumber\\
&\leq \Eb[X^2_i \cdot  \lv\{i\leq N(T)+1\}|\Gamma_i=\tau]\cdot\Eb[ \lv\{i\leq N(T)+1\}].
\end{align*}
\end{Lemma}
\begin{Proof}
Based on Cauchy-Schwarz inequality, we have
\begin{align}
&\left(\Eb[X_i \cdot \lv\{i\leq N(T)+1\}|\Gamma_i=\tau]\right)^2\nonumber\\
&\leq \Eb[X^2_i \cdot \left( \lv\{i\leq N(T)+1\}\right)^2|\Gamma_i=\tau]\\
&\quad\cdot\Eb[ \left(\lv\{i\leq N(T)+1\}\right)^2|\Gamma_i=\tau].\label{eqn:cauchy}
\end{align}
Lemma \ref{lemma:cauchy} then follows from the fact that $\lv\{i\leq N(T)+1\}$ is independent with $\Gamma_i$.
\end{Proof}

Last, we will show that the corresponding renewal policy always outperforms the original uniformly bounded policy in terms of AoI.
\begin{align}
& \Eb\left[\frac{R(S_{N(T)+1})}{T} \right]\nonumber\\
 &=\frac{1}{2T}\Eb\left[\sum_{i=1}^{\infty} X^2_i \cdot \lv\{i\leq N(T)+1\} \right]\\
 &=\frac{1}{2T}\sum_{i=1}^{\infty}\Eb[\Eb[X^2_i \cdot \lv\{i\leq N(T)+1\}|\Gamma_i=\tau]]\\
 &\geq \frac{1}{2T}\frac{\sum_{i=1}^{\infty}\Eb\left[\left(\Eb[X_i \cdot \lv\{i\leq N(T)+1\}|\Gamma_i=\tau]\right)^2\right]}{\Eb[\lv\{i\leq N(T)+1\}]}\label{eqn:holder}\\
 &= \frac{1}{2T} \sum_{i=1}^{\infty}\Eb\left[\left(\hat{X}^T_i(\tau)\right)^2\right]\Eb[\lv\{i\leq N(T)+1\}]\label{eqn:factorization2}\\
 &=\sum_{i=1}^{\infty}\Eb\left[\left(\hat{X}^T_i(\tau)\right)^2\right] \rho_i^T \cdot\frac{\Eb[N(T)+1]}{2T}\\
 &\geq \Eb_\tau \left[\left(\sum_{i=1}^\infty \hat{X}^T_i(\tau) \rho_i^T  \right)^2 \right]\cdot\frac{\Eb[N(T)+1]}{2T}\label{eqn:jensen}\\
 &=\Eb_\tau[\bar{X}^2_T(\tau)] \cdot\frac{\Eb[N(T)+1]}{2T} \geq \frac{\Eb_\tau[\bar{X}^2_T(\tau)]}{2\Eb_\tau[\bar{X}_T(\tau)]}\label{eqn:upper}\\
 &\geq \min_{X(\tau)\in \Pi'}\frac{\Eb_\tau[{X}^2(\tau)]}{2\Eb_\tau[{X}(\tau)]},\label{eqn:optimal}
\end{align}
where (\ref{eqn:holder}) follows from the Lemma \ref{lemma:cauchy}, (\ref{eqn:factorization2}) follows from Lemma~\ref{lemma:factorization}, (\ref{eqn:jensen}) follows from Jensen's inequality and Proposition~\ref{prop:pmf}. Combining with Lemma~\ref{lemma:first_moment}, we have (\ref{eqn:upper}), which is greater than or equal to (\ref{eqn:optimal}), the minimum long-term average AoI of the optimal renewal policy. We use $\Pi'$ to denote the set of feasible renewal policies under which $X_i$ only depends on $\Gamma_i$. Since the inequality holds for every $T$, we have
\begin{align}
\limsup_{T\rightarrow \infty}\Eb \left[ \frac{R(T)}{T}\right]&\geq \limsup_{T\rightarrow \infty} \Eb\left[\frac{R(S_{N(T)+1})}{T} \right]\\
&\geq  \min_{X(\tau)\in\Pi'}\frac{\Eb_\tau[{X}^2(\tau)]}{2\Eb_\tau[{X}(\tau)]}.
\end{align}

\subsection{Proof of Theorem~\ref{thm:renewal}}\label{appx:renewal}
Based on Theorem~\ref{thm:comparison}, we assume the inter-update delays under a renewal policy is a function of $\tau$, the duration between the last update epoch and the first energy arrival after it. 

Then, to minimize the long-term average AoI is equivalent to 
\begin{align}\label{opt_renewal}
\min_{X(\tau)}\quad \frac{\Eb_\tau [X^2(\tau)]}{2\Eb_\tau [X(\tau)]} \quad\mbox{s.t.}\quad X(\tau)\geq \tau, \forall \tau.
\end{align}
Based on the assumption that the energy arrival process is Poisson with $\lambda=1$, $\tau$ is an exponential random variable with rate 1.
%Since continuous function is dense in the space of measurable functions, it suffices to show that a continuous $X(\tau)$ is suboptimal to a threshold policy. 
\tcb{In order to make problem (\ref{opt_renewal}) more tractable to solve, we introduce the following parameterized problem
\begin{align}\label{opt_reform}
p(\lambda):= \min_{X(\tau)} \Eb_\tau [X^2(\tau)]-2\lambda \Eb_\tau [X(\tau)] \,\,\mbox{s.t.}\, X(\tau)\geq \tau, \forall \tau.
\end{align}}

We have the following observation.
\begin{Proposition}
The optimal solution of problem (\ref{opt_renewal}) is given by $\min\{\lambda^* :  \lambda^*\geq 0, p(\lambda^*)=0\}.$
\end{Proposition}

\tcb{Thus, in the following, we will first solve problem (\ref{opt_reform}) for any fixed $\lambda\geq 0$. We will then insert the obtained optimal solution for any $\lambda$ into (\ref{opt_reform}) and let it equal zero. The solution associated with the minimum $\lambda^*$ will be the optimal solution to problem (\ref{opt_renewal}). }

To solve (\ref{opt_reform}), we introduce the following Lagrangian \cite{boyd}:
\begin{align}
&L(X(\tau),\lambda,\mu(\tau))\nonumber\\
&=\int_{0}^{\infty} X^2(\tau) f_0(\tau)d\tau -2 \lambda \int_{0}^{\infty} X(\tau) f_0 (\tau)dt \nonumber\\
&\quad -\int_{0}^{\infty}\mu(\tau) [X(\tau)-\tau] f_0 (\tau)d\tau
\end{align}
where $\mu(\tau)$ is a non-negative Lagrange multiplier function, and $f_0(\tau)$ is the probability density function of $\tau$. Taking derivative with respect to $X(\tau)$ and setting it to zero, we have
\begin{align}\label{eqn:kkt}
 X(\tau)&=\lambda+\frac{\mu(\tau)}{f_0(\tau)},
\end{align}
and the complementary slackness condition indicates that
\begin{align}\label{eqn:slackness}
\mu(\tau) [X(\tau)-\tau]=0, \quad \forall \tau>0.
\end{align}
Thus, we have two possible cases:

Case 1) $X(\tau) \neq \tau$. In this case, we must have $\mu(\tau) = 0$ due to (\ref{eqn:slackness}). Plugging into (\ref{eqn:kkt}), we have $X(\tau) = \lambda $.

Case 2) $X(\tau) =\tau$. Due to the non-negativity of $\mu(\tau)$, and (\ref{eqn:kkt}), we have $\tau\geq \lambda$.

Combining both cases, $X(\tau)$ must be a function in the following form
 \begin{align}
X(\tau)&=\left\{\begin{array}{ll}
\lambda, & \tau\in(0,\lambda)\\
\tau, & \tau\geq \lambda
\end{array}\right.
\end{align}
which corresponds to a threshold policy.

Substituting this $X(\tau)$ into (\ref{opt_reform}), we have
\begin{align}
p(\lambda )=2e^{-\lambda} - \lambda^2
\end{align}
which admits a unique solution of  $\lambda^*= 0.9012$ when $p(\lambda )$ is equated to 0. This $\lambda^*$  also corresponds to the minimum AoI of the optimization problem in (\ref{opt_renewal}).
%%%%%%%%%

\end{document}